\documentclass[twocolumn]{aastex62}

\submitjournal{ApJ}

\shorttitle{Disrupting dwarf around NGC 2403}
\shortauthors{Carlin et al.}

\begin{document}

\title{Tidal destruction in a low mass galaxy environment: the discovery of tidal tails around DDO~44\footnote{Based on data collected at Subaru Telescope, which is operated by the National Astronomical Observatory of Japan.}}

\correspondingauthor{Jeffrey L. Carlin}
\email{jcarlin@lsst.org, jeffreylcarlin@gmail.com}

\author[0000-0002-3936-9628]{Jeffrey L. Carlin}
\affil{LSST, 950 North Cherry Avenue, Tucson, AZ 85719, USA} 

\author[0000-0001-9061-1697]{Christopher T. Garling}
\affil{CCAPP and Department of Astronomy, The Ohio State University, Columbus, OH 43210, USA}

\author[0000-0002-8040-6785]{Annika H. G. Peter}
\affil{CCAPP, Department of Physics, and Department of Astronomy, The Ohio State University, Columbus, OH 43210, USA}

\author{Denija Crnojevi\'c}
\affil{University of Tampa, 401 West Kennedy Boulevard, Tampa, FL 33606, USA}

\author{Duncan A. Forbes}
\affil{Centre for Astrophysics and Supercomputing, Swinburne University, Hawthorn VIC 3122, Australia}

\author{Jonathan R. Hargis}
\affil{Space Telescope Science Institute, 3700 San Martin Drive, Baltimore, MD 21218, USA}

\author{Bur\c{c}\.{i}n Mutlu-Pakd\.{i}l}
\affil{Department of Astronomy/Steward Observatory, 933 North Cherry Avenue, Rm. N204, Tucson, AZ 85721-0065, USA}

\author{Ragadeepika Pucha}
\affil{Department of Astronomy/Steward Observatory, 933 North Cherry Avenue, Rm. N204, Tucson, AZ 85721-0065, USA}

\author{Aaron J. Romanowsky}
\affil{University of California Observatories, 1156 High Street, Santa Cruz, CA 95064, USA}
\affil{Department of Physics \& Astronomy, San Jos\'e State University, One Washington Square, San Jose, CA 95192, USA}

\author[0000-0003-4102-380X]{David J. Sand}
\affil{Department of Astronomy/Steward Observatory, 933 North Cherry Avenue, Rm. N204, Tucson, AZ 85721-0065, USA}

\author{Kristine Spekkens}
\affil{Department of Physics, Engineering Physics and Astronomy, Queen's University, Kingston, Ontario, Canada, K7L 3N6}
\affil{Department of Physics, Royal Military College of Canada, P.O. Box 17000, Station Forces, Kingston, ON K7L 7B4, Canada}

\author{Jay Strader}
\affil{Department of Physics and Astronomy, Michigan State University, East Lansing, MI 48824, USA}

\author{Beth Willman}
\affil{Department of Astronomy/Steward Observatory, 933 North Cherry Avenue, Rm. N204, Tucson, AZ 85721-0065, USA}
\affil{Association of Universities for Research in Astronomy, 950 North Cherry Avenue, Tucson, AZ 85719, USA}

\begin{abstract}

We report the discovery of a $\gtrsim1\arcdeg$ ($\sim50$~kpc) long stellar tidal stream emanating from the dwarf galaxy DDO~44, a likely satellite of Local Volume galaxy NGC~2403 located $\sim70$~kpc in projection from its companion. NGC~2403 is a roughly Large Magellanic Cloud stellar-mass galaxy 3 Mpc away, residing at the outer limits of the M~81 group. We are mapping a large region around NGC~2403 as part of our MADCASH (Magellanic Analogs' Dwarf Companions and Stellar Halos) survey, reaching point source depths (90\% completeness) of ($g, i$) = (26.5, 26.2). Density maps of old, metal-poor RGB stars reveal tidal streams extending on two sides of DDO~44, with the streams directed toward NGC~2403. We estimate total luminosities of the original DDO~44 system (dwarf and streams combined) to be $M_{i, \rm{tot}} = -13.4$ and $M_{g, \rm{tot}} = -12.6$, with $\sim25-30\%$ of the luminosity in the streams. Analogs of $\sim$LMC-mass hosts with massive tidally disrupting satellites are rare in the Illustris simulations, especially at large separations such as that of DDO~44. The few analogs that are present in the models suggest that even low-mass hosts can efficiently quench their massive satellites.

\end{abstract}

\keywords{galaxies: dwarf, galaxies: halos, galaxies: individual (NGC~2403, DDO~44), galaxies: photometry}


\section{Introduction} \label{sec:intro}

Deep surveys covering large sky areas have in recent years greatly expanded the number of dwarf galaxy satellites known around the Milky Way \citep[MW; e.g.,][]{Kim2015,Drlica-Wagner2015,Drlica-Wagner2016,Laevens2015,Homma2018,Torrealba2016,Torrealba2018} and M31 \citep[e.g.,][]{Martin2016a, McConnachie2018}. The newly discovered diminutive galaxies include extremely low-luminosity ``ultra-faint dwarfs'' (UFDs; e.g., \citealt{Willman2005, Belokurov2007b} --- see the recent review by \citealt{Simon2019}), as well as many relic streams from tidally disrupted satellites criss-crossing the Galactic halo (e.g., \citealt{Belokurov2006a,Grillmair2016,Shipp2018}). In parallel to these discoveries, models of structure formation and evolution within the $\Lambda$-Cold Dark Matter ($\Lambda$CDM) framework have generated predictions of the number of satellites expected, as well as properties such as their luminosity functions and metallicities \citep{Benson2002,Zolotov2012,Wetzel2016,Jethwa2018,Bose2018,Kim2018,Nadler2019}. Although the match with the MW and M31 --- the only systems with robust samples of satellites --- is remarkably good, it is unclear whether our Galaxy and its nearest massive neighbor are representative of massive galaxies more generally, or if theoretical models are over-tuned to the Local Group. 
Resolved stellar maps of nearby massive galaxies are now being painstakingly assembled, revealing the satellite systems (down to the scale of ultrafaint dwarf galaxies) of Cen A \citep{Crnojevic2019}, NGC~253 \citep{Sand2014, Romanowsky2016, Toloba2016}, M81 \citep{Chiboucas2013}, M101 \citep{Merritt2014, Bennet2017,Danieli2017, Muller2017,Bennet2019} and M94 \citep{Smercina2018}, among others. We are thus entering an era in which we may explore the stochasticity of satellite populations around a variety of hosts, as well as their dependence on environment and host properties \citep[e.g.][]{Bennet2019}.  These can be used to make more precise tests of galaxy formation and the $\Lambda$CDM cosmological model.

Of particular interest are satellites in less dense environments than the ones highlighted above.  Satellites of the MW in particular show signs of experiencing  many types of environmental quenching and disruption simultaneously \citep{barkana1999,mayer2006,Grcevich2009,nichols2011,brown2014,slater2014,fillingham2015,wetzel2015}.  Because so many processes are likely to affect the satellites, it is often difficult to assess their relative importance, and how that importance scales with properties of the host (not necessarily limited to halo mass). One way to disentangle these processes, and to highlight the scales at which each mechanism kicks in, is to consider low-mass hosts.  Hosts inhabiting halos smaller than the Milky Way's ought not to have hot-gas halos \citep{birnboim2003}, so ram-pressure stripping and possibly starvation may be significantly reduced as compared to the MW \citep[although they are likely to have cool circumgalactic media (CGM);][]{bordoloi2014}.  Moreover, they should have gentler tidal fields, reducing the effects of tidal heating and stripping.  Thus, we expect satellite galaxies of low-mass hosts to be more like field galaxies, and less influenced by their environment. The environmental processes that are important for less massive hosts are likely to be different than those most relevant to MW-sized galaxies.  These hypotheses remain to be tested.

We have an additional motivation to study satellites of low-mass galaxies, in that many of the recently discovered dwarf galaxies within the MW halo are thought to have originated as satellites of the Large and Small Magellanic Clouds (LMC, SMC), and only recently fell into the MW \citep[e.g.,][]{Jethwa2016,Sales2017,Dooley2017b,kallivayalil2018}.  The luminosity function of the new discoveries is unexpected, though --- there are no massive ($M_* > 10^4 M_\odot$) MC satellite candidates (though \citealt{Pardy2019} suggest that the Carina and Fornax dSphs may be associated with the MCs), but many that are much smaller, at odds with typical stellar-mass--halo-mass relations \citep{Dooley2017b}.\footnote{A similar result is found for the MW as a whole \citep{Kim2018}.}  It is unknown whether the LMC and SMC had a more typical luminosity function at infall and many satellites have been stripped from them by the MW, or if this luminosity function is typical and is telling us something new about galaxy formation in small halos.

To extend the mass range of hosts for which satellite searches have been carried out to lower-mass (Magellanic Cloud-mass) systems, without the difficulty of interpreting the interplay of the LMC and its satellites with the Galactic halo, we are conducting a census of nearby LMC stellar-mass analogs. With this survey --- Magellanic Analogs' Dwarf Companions and Stellar Halos (MADCASH) --- we are searching for the satellite populations of MC-mass galaxies within $\sim4$~Mpc of the MW.
Some early results from this ongoing survey include the discovery of the $M_{\rm V} \sim -9.7$ dwarf galaxy Antlia~B near NGC~3109 \citep{Sand2015, Hargis2019}, the detection of extended stellar populations around nearby galaxy IC~1613 \citep{Pucha2019}, and our discovery of a faint ($M_{\rm V} \sim -7.7$) satellite of NGC~2403 \citep{Carlin2016}.
In this work, we highlight the discovery of a dwarf satellite being tidally disrupted around nearby ($D \sim 3.2$~Mpc; \citealt{Karachentsev2013}) low-mass (stellar mass $M_{*} \sim 7\times10^9~M_\odot$; roughly $2\times$ LMC stellar mass) spiral galaxy NGC~2403, a relatively isolated system at the outskirts of the M81 group.

The dwarf spheroidal DDO~44 is a relatively massive dwarf ($M_{\rm R} \sim -13.1$, similar to the Fornax satellite of the MW; \citealt{Karachentsev1999a}) that is at a distance and velocity consistent with orbiting as a satellite of NGC~2403. Here we report evidence that the dwarf spheroidal DDO~44 has stellar tidal tails extending at least $\sim0.5\arcdeg$ ($\sim25$~kpc) from its center. This discovery is based on data from our deep, wide-area imaging survey to a projected radius of $\sim100$~kpc around NGC~2403.  

In Sec. \ref{sec:data}, we introduce our discovery data set and analysis procedure.  We show the key characteristics of the stream in Sec. \ref{sec:stream}.  In Sec. \ref{sec:context}, we discuss what the stream implies for the relationship between the orbital and star-formation histories of DDO 44, and the frequency of small galaxy disruption by low-mass hosts.  We highlight our key results in Sec. \ref{sec:discussion}.

\begin{figure}[!t]
\includegraphics[width=0.95\columnwidth, trim=0.0in 0.0in 0.75in 0.5in, clip]{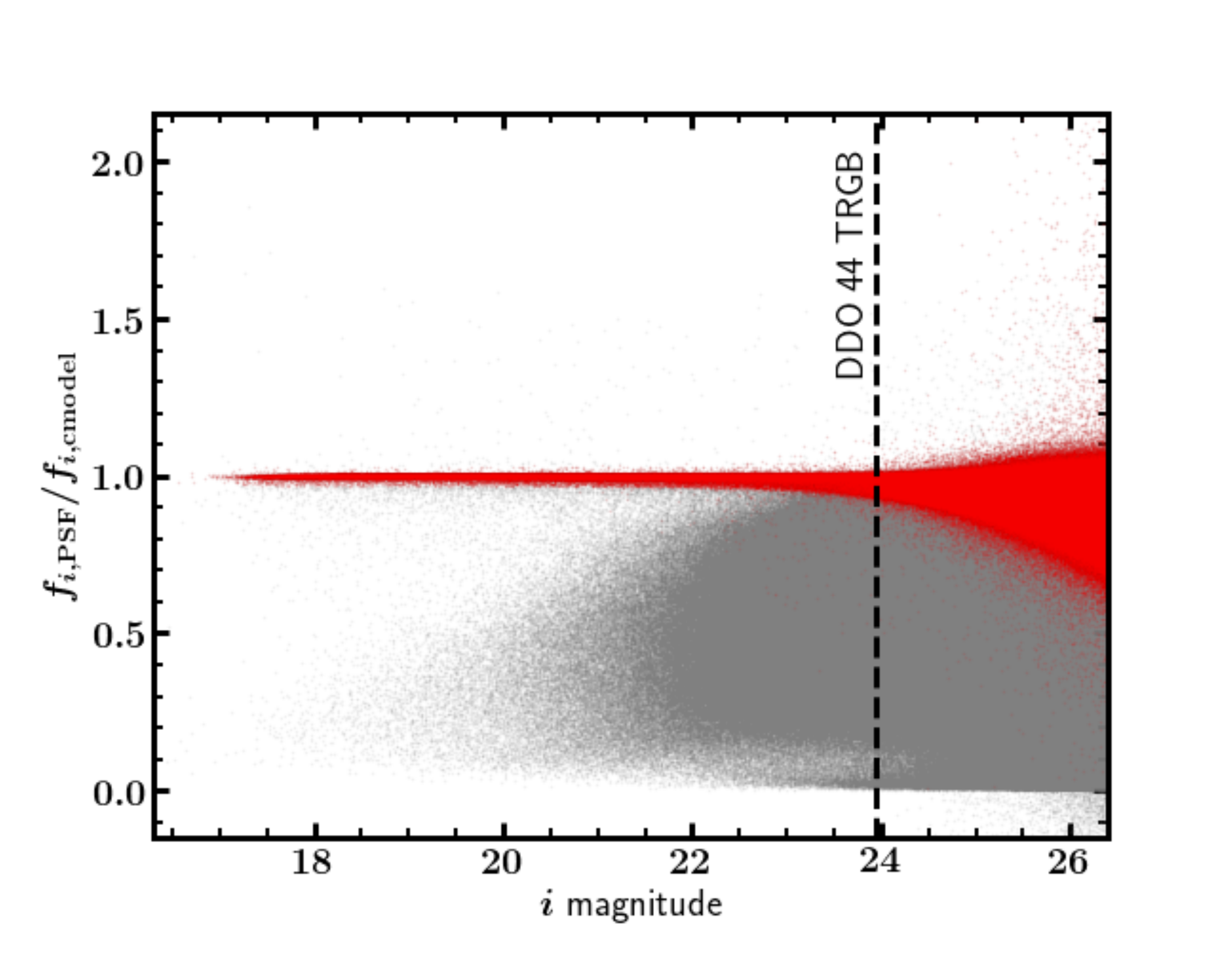}
\caption{Star/galaxy separation based on the ratio of PSF to \emph{cmodel} fluxes. The measured flux ratios of sources, $f_{i, \rm PSF}/f_{i, cmodel}$, are shown as a function of $i$-band PSF magnitude; flux ratios of $\sim1$ correspond to point-like sources. We classify all objects whose flux ratios are consistent (within their measured $1\sigma$ errors) with $f_{i, \rm PSF}/f_{i, cmodel} \sim 1\pm0.03$ as point sources (i.e., ``stars''). These are shown as red points above, with all other excluded sources shaded gray. For reference, our measured magnitude of the DDO~44 tip of the red giant branch (TRGB) is plotted as a dashed vertical line. \label{fig:star_gx}}
\end{figure}

\section{Data and Analysis} \label{sec:data}

Deep imaging data were obtained with Hyper Suprime-Cam (HSC; \citealt{Furusawa2018, Kawanomoto2018, Komiyama2018, Miyazaki2018}) on the Subaru 8.2m telescope. The $1.5^\circ$ diameter field of view of HSC corresponds to $\sim80$~kpc at the $D\sim3.0$~Mpc distance of NGC~2403, enabling a relatively efficient survey to beyond a projected radius of $d > 100$~kpc around NGC~2403 (a large fraction of the $\sim120-180$~kpc virial radius of an isolated LMC-mass analog; see, e.g., estimates of $R_{\rm vir}$ in \citealt{Dooley2017b}). Our data consist of seven HSC pointings (see map in Figure~\ref{fig:rgb_density}): the CENTER, EAST, and WEST fields were observed on 2016 February 9--10, while the four additional HSC fields (NW, NE, SW, and SE) were observed 23--24 December 2017. All observations consist of $10\times300$s exposures in $g$-band (known as ``HSC-G'' at Subaru) and $10\times120$s in $i$ (``HSC-I2''). We also observed short, $5\times30$s sequences of exposures to improve photometry at the bright end. All observations from both runs were obtained in seeing between $\sim0.5-0.9''$, under clear skies.

The data were processed with the LSST pipeline, a version of which was forked to create the reduction pipeline for the HSC-SSP survey \citep{Aihara2018a, Aihara2018b}. Details of the reduction steps can be found in \citet{Bosch2018}. In short, we performed forced PSF photometry on co-added frames in each filter, and calibrated both astrometrically and photometrically to PanSTARRS-1 (PS1; \citealt{Schlafly2012, Tonry2012, Magnier2013}). We applied extinction corrections based on the \citet{Schlafly2011} coefficients derived from the \citet{Schlegel1998} dust maps. All results presented in this work are based on extinction-corrected PSF magnitudes.

\begin{figure}[!t]
\includegraphics[width=0.95\columnwidth]{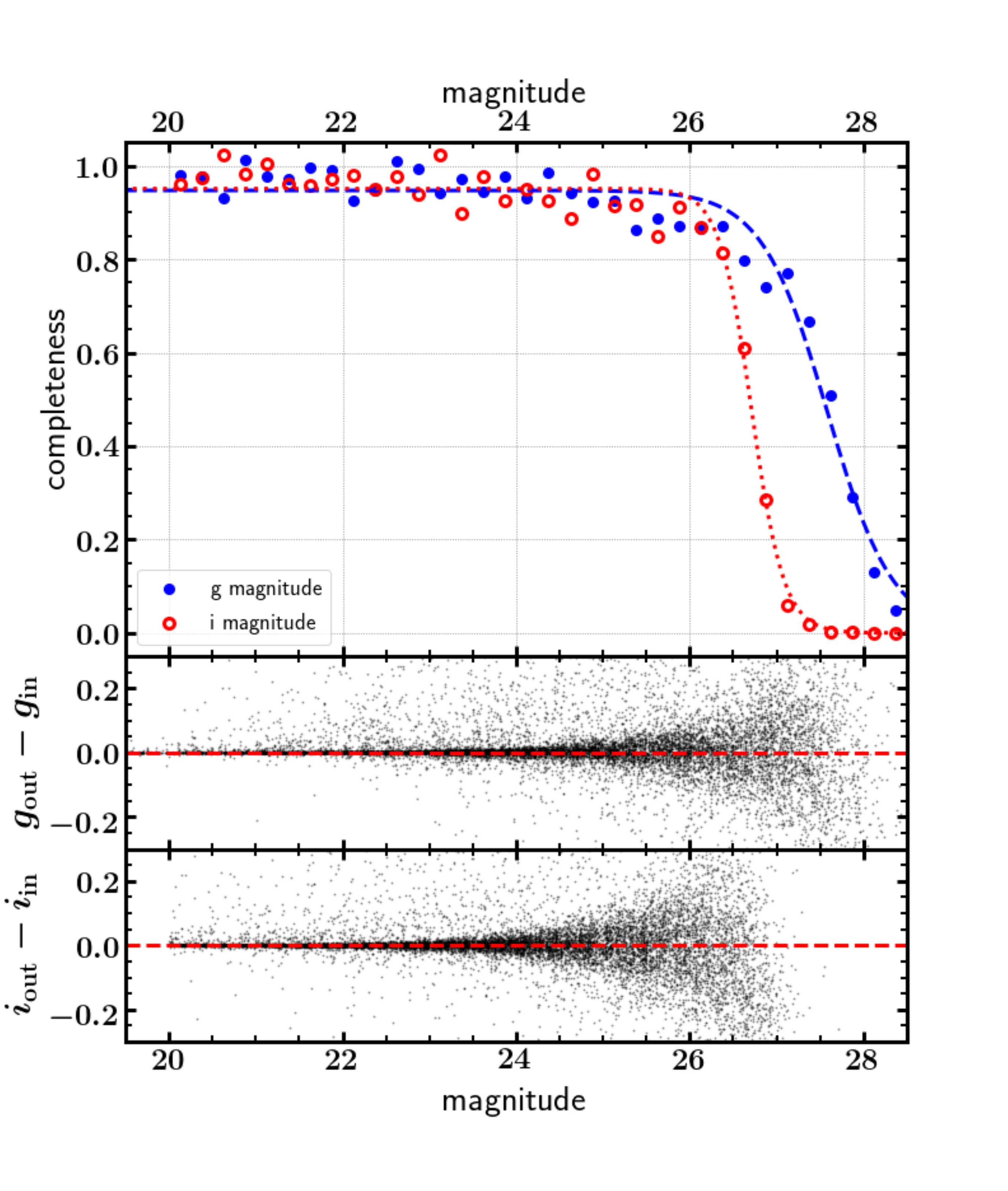}
\caption{Photometric completeness (top panel) as a function of magnitude in the $20\arcmin$ region around DDO~44. Dashed/dotted lines show fits to the completeness curves of the form used in \citet{Martin2016a}; the data are 50\% (90\%) complete for point sources at $i = 26.7 (26.2)$ and $g = 27.5 (26.5)$. The lower panels compare recovered and input artificial star magnitudes, demonstrating that no systematic error is present in our photometric measurements. \label{fig:completeness}}
\end{figure}

For separation of point sources from unresolved background galaxies, we compare the ratio of PSF to \emph{cmodel} fluxes for all sources, where the \emph{cmodel} is a composite bulge/exponential plus de Vaucouleurs profile fit to each source. Point sources should have flux ratios $f_{\rm PSF}/f_{\rm cmodel} \sim 1$, while extended sources will contain additional flux in the model measurement that is not captured by the PSF. Figure~\ref{fig:star_gx} shows the $i$-band flux ratio as a function of $i$-band PSF magnitude. A large number of sources (especially at the bright end) are concentrated around unity in this figure. We allow for an intrinsic width of $\pm0.03$ in the flux ratio, and select sources whose $1\sigma$ uncertainties in flux ratio place them within this $\pm0.03$ window. The point source candidates selected in this way are shown as red points in Figure~\ref{fig:star_gx}, with extended sources (i.e., ``not point sources'') as gray points. Note that some background galaxies will contaminate the point source sample below $i\sim23$, where background galaxies far outnumber stars. Through the rest of this work we will analyze only this point source sample calibrated to the PS1 photometric system.

\begin{figure*}[!t]
\begin{center}
\includegraphics[width=0.85\textwidth]{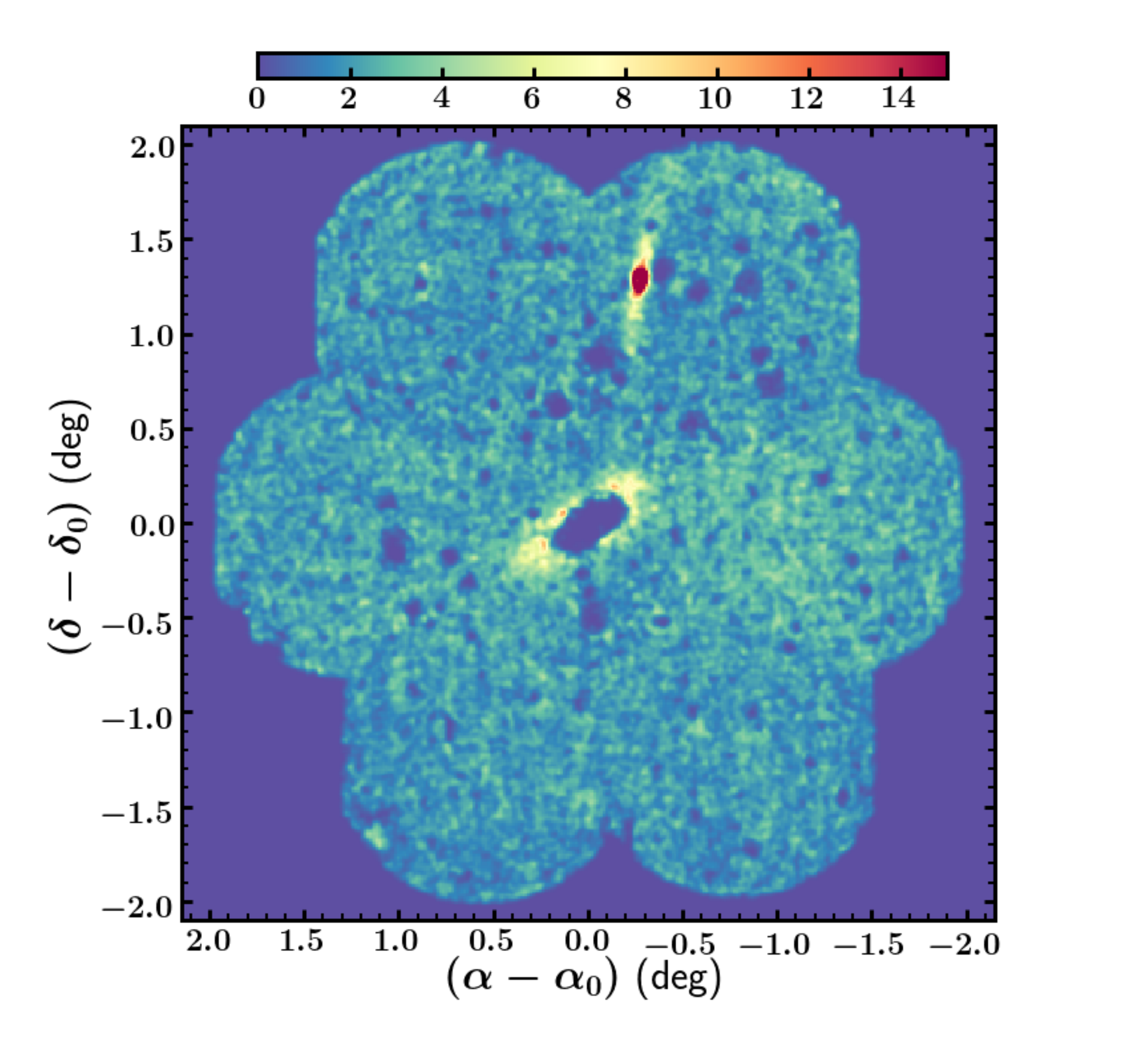}
\caption{Density map of candidate RGB stars at the distance and metallicity of DDO~44/NGC~2403 (selected using the RGB box in Figure~\ref{fig:cmd_ddo44}). Bins are $0.75\arcmin$, and the image has been smoothed with a Gaussian kernel of $0.75\arcmin$ FWHM. The field is centered on NGC~2403 (the hole in the center is due to extreme crowding), and DDO~44 is to the north (and slightly west). NGC~2366 is $\sim2.4\arcdeg$ north of DDO~44. Stellar number densities have been corrected for completeness as a function of position; the colorbar encodes the number of stars per $0.75\arcmin$ bin. North is up  and East is to the left.  \label{fig:rgb_density}}
\end{center}
\end{figure*}

To characterize the completeness of our photometric catalog, we injected artificial stars into the images using \texttt{Synpipe} \citep{Huang2018}, which was written for HSC-SSP, and has been incorporated into the LSST pipeline. The resulting completeness (i.e., the fraction of artificial stars recovered by the photometric pipeline) in a $20\arcmin$ region centered on DDO~44 (excluding the central $2\arcmin$ where crowding is too extreme to resolve stars; a total of 18975 artificial stars were injected in this region, or $\sim15~{\rm arcmin}^{-2}$) is given in Figure~\ref{fig:completeness}. We fit a function of the form given in \citet{Martin2016a} to these curves, and estimate 50\% (90\%) completeness limits of $i = 26.7~(26.2)$ and $g = 27.5~(26.5)$. In the lower panels of the figure, we compare the input and recovered magnitudes for the artificial stars in both bands. These are centered on zero, so we are confident that no systematic offsets are present in our photometry.

\begin{figure*}[!t]
\begin{center}
\includegraphics[width=0.475\textwidth]{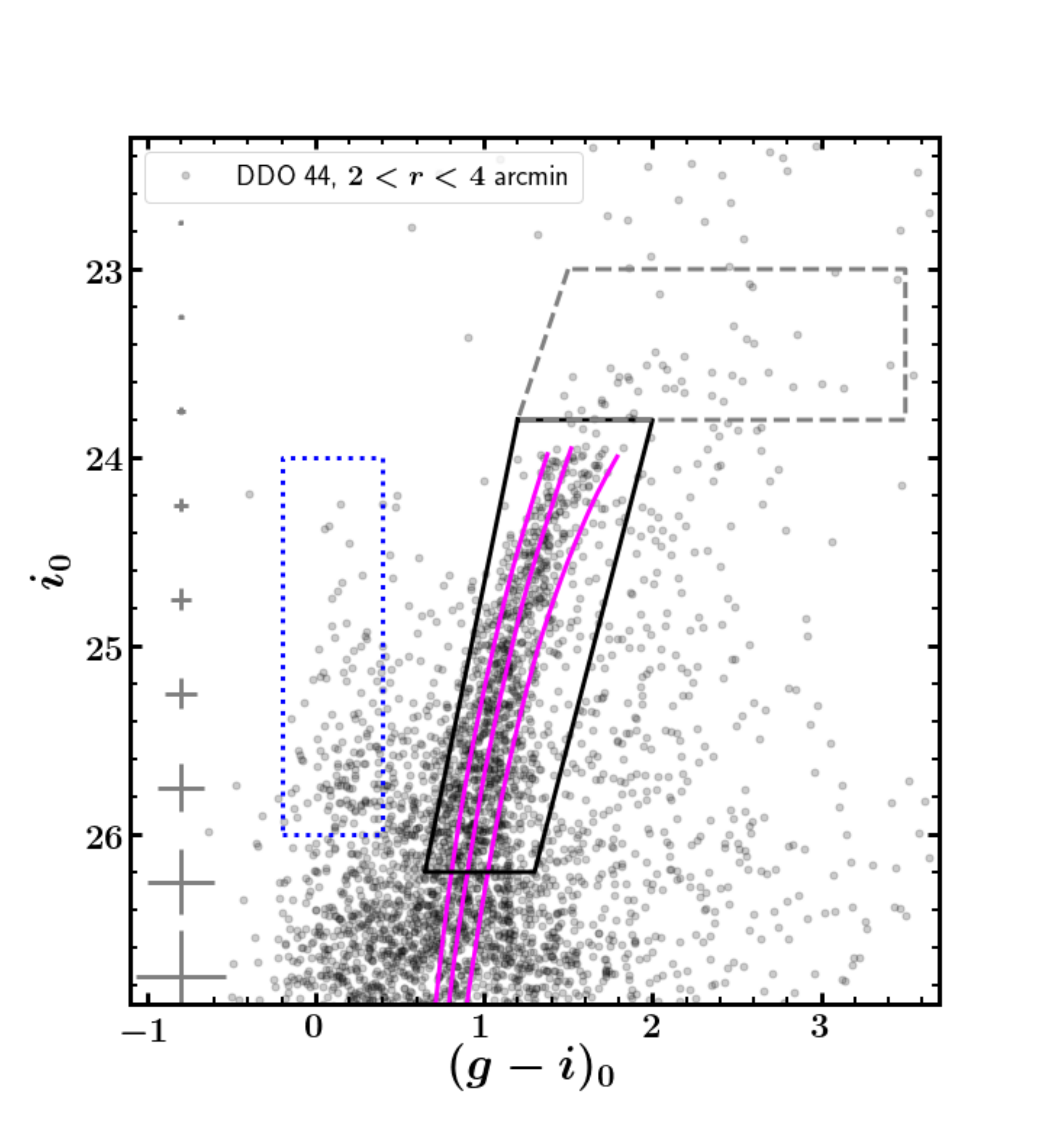}
\includegraphics[width=0.475\textwidth]{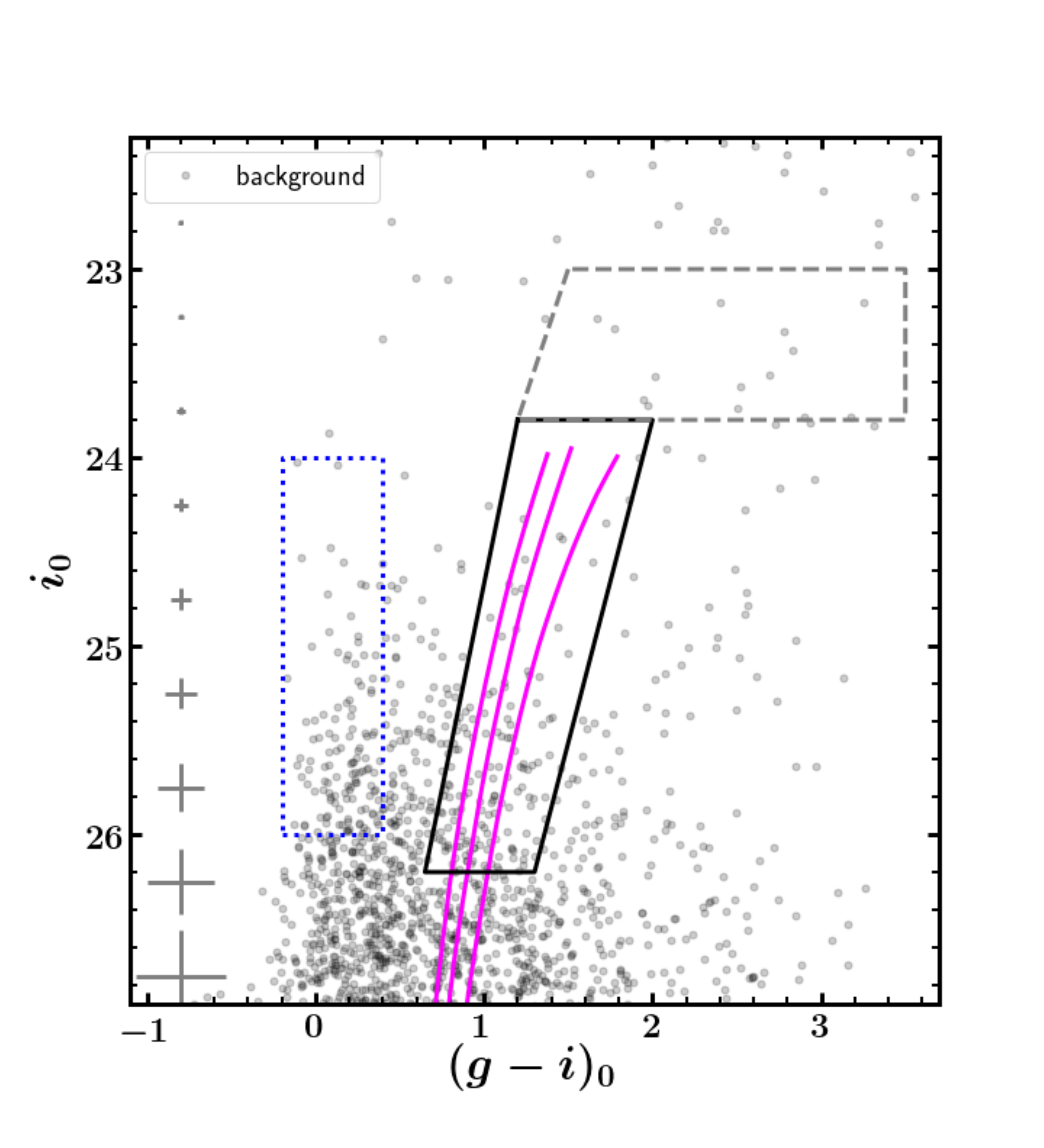}
\caption{{\it Left panel:} Color magnitude diagram of stars between $2-4\arcmin$ of the center of DDO~44, showing a clearly defined old, metal-poor RGB of DDO~44, with few (if any) young stars. We exclude the region inside $r < 2\arcmin$ because the photometry in this inner region is affected by crowding; blending and elevated background due to unresolved stars is also the source of the large scatter (especially redward of the RGB) in the DDO~44 CMD. We overlay MIST isochrones \citep{Choi2016,Dotter2016a} for old (10~Gyr) populations with [Fe/H] = $-1.6$ (our estimate for the metallicity of DDO~44; see Section~\ref{sec:dist_isofit}) flanked by isochrones with metallicities $\pm0.5$ dex, and shifted to our measured distance modulus of 27.36. The black box shows the selection used to isolate candidate RGB stars for all analysis in this paper. The other polygons outline regions used to select candidate AGB (dashed gray box) and blue sources (dotted blue outline) for Figure~\ref{fig:stream_density}. Median photometric uncertainties as a function of magnitude are shown near the left edge of the plot. {\it Right panel:} CMD of sources in a region shifted southwest by $0.2^\circ$ in both RA and Dec, but of the same size as the DDO~44 field. This highlights (a) the lack of stars within the RGB box relative to the left panel, and (b) a similar population of sources within the blue box as seen in the DDO~44 CMD, suggesting that these are unresolved background galaxies rather than blue stars associated with DDO~44. 
\label{fig:cmd_ddo44}}
\end{center}
\end{figure*}

\section{A Stream around DDO~44} \label{sec:stream}

One of the primary goals of our large-area imaging campaign around NGC~2403 is to search for its dwarf galaxy companions and/or the remnants of destroyed satellites. Thus one of the first things we did upon finishing the data reduction was to select stars with color-magnitude diagram (CMD) positions consistent with metal-poor RGB stars at the distance of NGC~2403, and plot their density on the sky. This RGB density map centered on NGC~2403 is shown in Figure~\ref{fig:rgb_density}. For the most part, the map shows a fairly uniform distribution over the entire region surveyed. This is most likely predominantly fore-/back-ground contamination, with little or no ``halo'' population in the outer regions around NGC~2403. We note that the $0.75'$ bins in this map are about the size of a typical faint dwarf galaxy at the distance of NGC~2403 --- $1' \approx 0.9$~kpc at $D = 3.0$~Mpc. 
Thus the faint dwarf galaxy MADCASH J074238+652501-dw found by \citet{Carlin2016} is almost completely contained in a single pixel of this map (approximately at $(\alpha-\alpha_0)\sim0.5$ and $(\delta-\delta_0)\sim-0.2$), and thus not visible as an obvious overdensity. The most striking feature in Figure~\ref{fig:rgb_density} is the prominent blob corresponding to the known dwarf spheroidal galaxy DDO~44 to the north (and slightly west) of NGC~2403. Our deep Subaru+HSC data enable us to see for the first time that DDO~44 has streams of stars emanating from it, oriented along the direction toward (and away from) NGC~2403; i.e, DDO~44 is tidally disrupting beyond doubt.

\begin{figure*}[!t]
\includegraphics[width=0.57\textwidth]{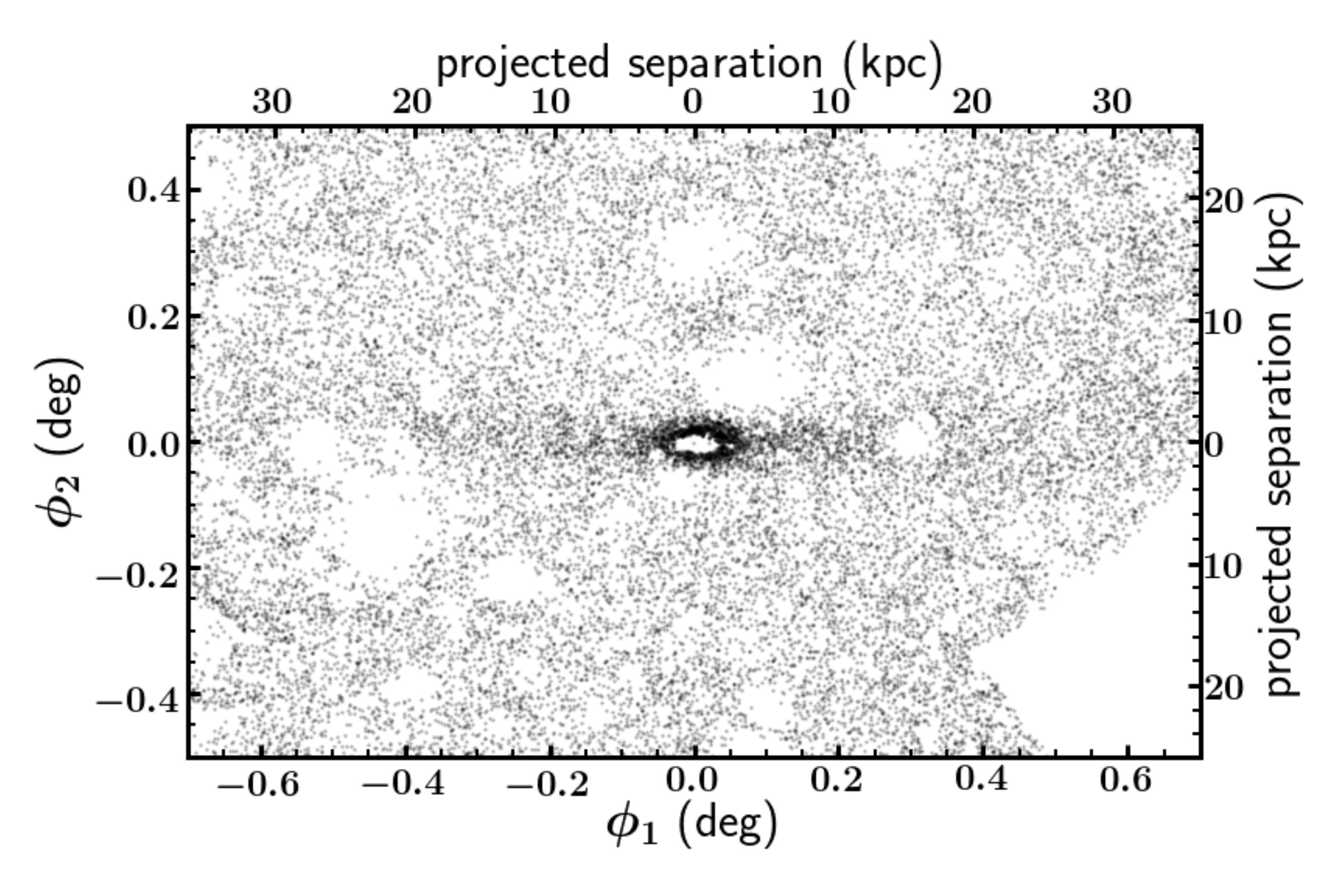}
\includegraphics[width=0.42\textwidth, trim=0in 0.5in 0in 0.5in, clip]{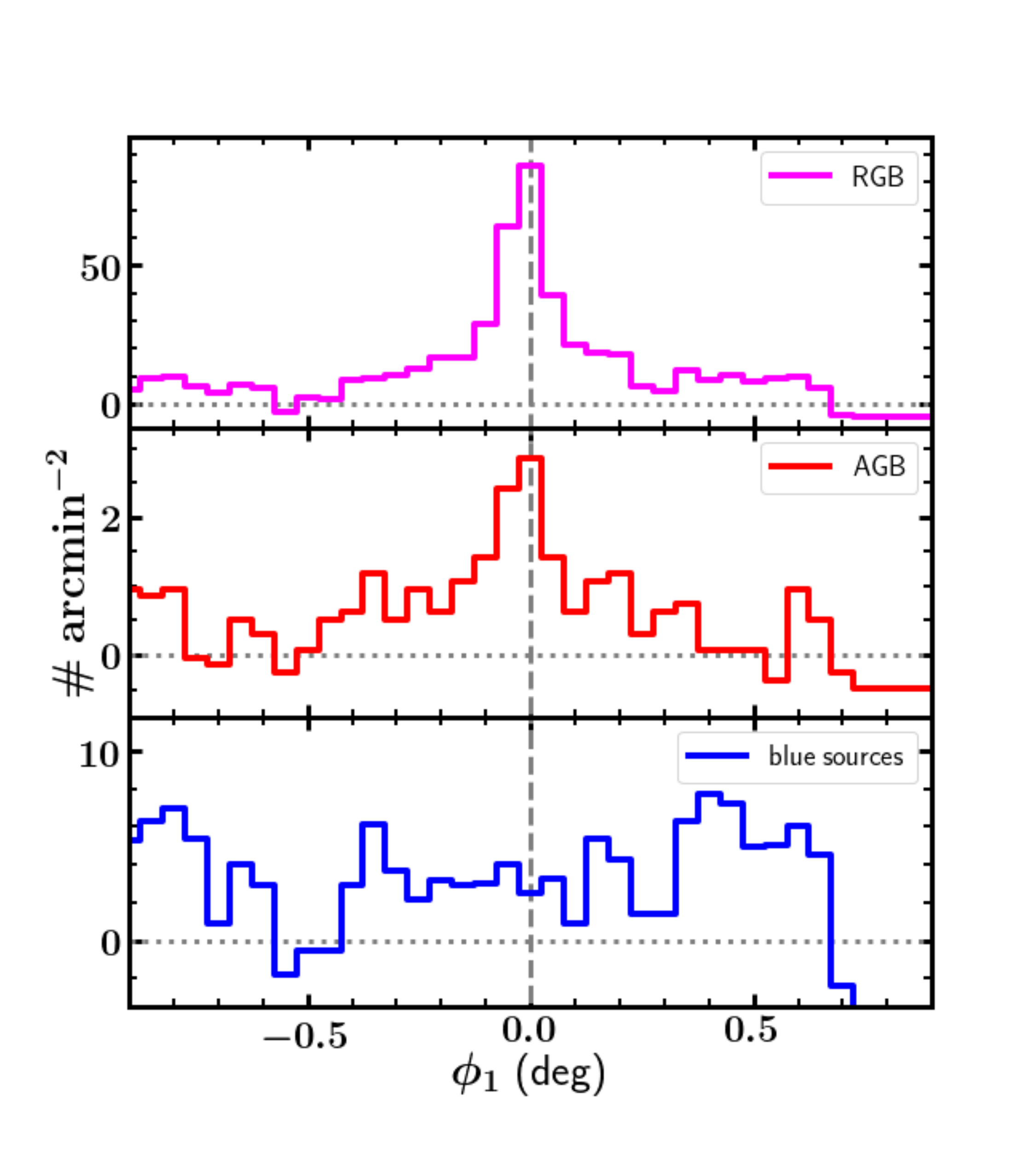}
\caption{{\it Left panel:} Scatter plot showing positions (in coordinates aligned with the DDO~44 stream) of RGB candidate stars, showing the clear detection of the stream as an overdensity. The upper and right axes show the projected separation in kpc assuming a distance of 2.96~Mpc (Section~\ref{sec:dist_isofit}). Empty circular regions are ``halos'' around bright stars where the detection algorithm flags sources as unreliable. {\it Right panel:} Number density of sources along the stream in a strip within $\phi_2 \pm 0.05\arcdeg$. From top to bottom, the panels show sources selected using the polygons in Figure~\ref{fig:cmd_ddo44} to highlight RGB stars (top panel), AGB stars (middle panel), and blue sources (bottom panel). In each panel, we have subtracted off the mean background in surrounding regions that are free from ``holes'' due to bright stars. There is a peak in the AGB profile at $\phi_1 \sim 0\arcdeg$ corresponding to the RGB peak. In the lower panel, no corresponding peak is seen among the blue sources, suggesting that these are unresolved background galaxies rather than young stars associated with DDO~44. \label{fig:stream_density}}
\end{figure*}

\subsection{TRGB distance, isochrone fit}\label{sec:dist_isofit}

In Figure~\ref{fig:cmd_ddo44} (left panel) we show a CMD of all stars between 2--4~arcmin of the center of DDO~44. 
There is a prominent metal-poor RGB (highlighted by the solid black box) as well as a significant number of AGB stars (dashed gray box) signaling the presence of intermediate-age ($\sim2-8$~Gyr old) stellar populations (as was found in \textit{HST} imaging by \citealt{Karachentsev1999a} and \citealt{Alonso-Garcia2006}). The dotted blue box in each panel denotes the location where young main-sequence stars would appear, if present. By comparing the DDO~44 field with an equal-area background region (right panel), the number of objects in this box is consistent with being background (likely unresolved galaxies/QSOs, given their blue colors). In our later analysis, we show (e.g., the lower panel of Fig.~\ref{fig:stream_density}) that there is no concentration of blue sources at the position of DDO~44, confirming that these are background objects rather than young stars in DDO~44.

To refine the distance of DDO~44, we first selected stars with colors consistent with metal-poor RGB stars ($0.8 < (g-i)_0 < 2.1$), and within $6'$ of the center of DDO~44. We binned these stars as a function of magnitude, then convolved the luminosity function with a zero-sum Sobel edge-detection filter (in particular, one with values [$-1.0, -2.0, -1.0, 0.0, 1.0, 2.0, 1.0$], as in \citealt{Jang2017}). We identify a narrow peak in the convolved luminosity function, corresponding to the tip of the RGB (TRGB), at $i_{\rm TRGB} = 23.95\pm0.05$. From 10-Gyr Dartmouth isochrones \citep{Dotter2008} at [Fe/H]$ = -1.6\pm0.3$, we estimate an $i$-band TRGB absolute magnitude\footnote{Note that the $i$-band TRGB magnitude is virtually independent of metallicity for metal-poor populations (as we confirmed with isochrones of various metallicities), so that our choice of metallicity has no bearing on the derived distance modulus.} (in the PS1 system) of $-3.41^{+0.02}_{-0.01}$. We thus derive a distance to DDO~44 of $2.96\pm0.10$ Mpc (i.e., distance modulus $(m-M)_0 = 27.36\pm0.07$). We note that the same TRGB code applied to the main body of NGC~2403 yields an identical distance modulus of $(m-M)_{\rm N2403} = 27.36$.
Our derived distance modulus is in agreement with other recent determinations for DDO~44 ($(m-M)_0 = 27.36\pm0.09, 27.39\pm0.13$, and 27.45; \citealt{Jacobs2009, Alonso-Garcia2006, Dalcanton2009}), though somewhat at odds with the value of $(m-M)_0 = 27.50$ given in the COSMICFLOWS-3 database \citep{Tully2016}.

At a distance of 2.96~Mpc, the on-sky separation between DDO~44 and NGC~2403 of $78.3\arcmin$ corresponds to a projected separation of $\sim67$~kpc. 

After determining the TRGB magnitude (and thus distance) of DDO~44, we wish to estimate the system's metallicity. To do so, we create a set of old (10~Gyr) Dartmouth isochrones in 0.1-dex metallicity intervals, and use a least-squares minimization based on differences between the DDO~44 stellar sample and the isochrones to find a best-fitting metallicity of [Fe/H]$ = -1.6\pm0.3$. This mean metallicity is consistent with those measured by \citet[][$-1.7\pm0.4$]{Karachentsev1999a}, \citet[][$-1.54\pm0.14$]{Alonso-Garcia2006}, and \citet[][$-1.67\pm0.19$]{Lianou2010} via \textit{HST} imaging. Figure~\ref{fig:cmd_ddo44} shows a CMD of the central $2-4\arcmin$ field around DDO~44, with the best fit isochrone at [Fe/H] = $-1.6$ overlaid, along with isochrones at $\pm0.5$~dex in metallicity. 

The RGB of DDO~44 is wider than expected solely based on photometric errors. 
This could be due in part to photometric scatter induced by the significant unresolved emission in the body of DDO~44. However, \citet{Alonso-Garcia2006} estimated that as much as $20\%$ of the total stellar content of DDO~44 is contributed by the intermediate-age population (likewise, \citealt{Lianou2010} found the fraction of AGB stars relative to RGB number to be $f_{\rm AGB} = 0.11$). Thus, a single 10-Gyr population should not be expected to reproduce the width of the RGB. Estimating the relative contributions of the different age populations (i.e., a star formation history) is beyond the scope of the current study (and is typically best achieved with data reaching the oldest MSTO). Finally, we note that we calculated a metallicity distribution under the assumption that only 10 Gyr populations were present (assigning stellar metallicities based on isochrones), and found a metallicity spread of $\sim0.49$~dex (determined by fitting a Gaussian to the distribution). The mean metallicity and metallicity spread (with the caveat that we have assumed a single age) is similar to those of Milky Way dSphs with similar luminosities (e.g., the Sculptor dSph; \citealt{Simon2019}), suggesting that some of the RGB width is contributed by a metallicity spread in DDO~44, while the presence of intermediate-age populations may account for some additional broadening of the RGB. 

Based on the lack of blue stars in \textit{HST} images of DDO~44, \citet{Karachentsev1999a} estimated that the most recent star formation in DDO~44 was at least 300~Myr ago. We do not see evidence of this young population beyond $2\arcmin$. However, a significant population of bright AGB stars above the TRGB (also seen in \textit{HST} data by \citealt{Karachentsev1999a} and \citealt{Alonso-Garcia2006}) suggests that intermediate-age populations are present in the outer regions of DDO~44. Indeed, as noted previously, \citet{Alonso-Garcia2006} found that $\sim20\%$ of the stellar population of DDO~44 consists of intermediate-age (between $\sim5-8$ Gyr, and at least older than 2 Gyr) populations.

\subsection{Stellar populations in the stream}\label{sec:stream_pops}

To facilitate analysis of the stream, we first derived the transformation to a coordinate system aligned with the stream. We determined the central position at points along the stream by fitting Gaussians to the stellar density in slices of $0.1^\circ$ in declination. Using the two points immediately adjacent to DDO~44, but on the north and south side, we derived the transformation to a great circle coordinate frame using the \texttt{gala}\footnote{\url{http://gala.adrian.pw/en/latest/}} software. This transformation places DDO~44 at the origin, with angle $\phi_1$ along the stream, and $\phi_2$ perpendicular to the stream. A map of RGB stars in the transformed coordinates is shown in the left panel of Figure~\ref{fig:stream_density}.

We then selected narrow strips of $\left|\phi_2\right| < 0.05^\circ$, and extracted an RGB star density profile as a function of $\phi_1$ (i.e., along the stream). This profile is shown in the right panel of Figure~\ref{fig:stream_density}, where we have subtracted the mean density in background regions that do not contain bright star holes (seen as white voids in Fig.~\ref{fig:stream_density}). The three background regions are at $-0.7\arcdeg < \phi_1 < -0.1\arcdeg$, $0.15\arcdeg < \phi_2 < 0.4\arcdeg$; $-0.2\arcdeg < \phi_1 < 0.35\arcdeg$, $-0.4\arcdeg < \phi_2 < -0.15\arcdeg$; and $0.2\arcdeg < \phi_1 < 0.7\arcdeg$, $0.15\arcdeg < \phi_2 < 0.4\arcdeg$.
This density profile shows RGB overdensities extending to at least $0.3^\circ$ from DDO~44 on either side, or $\sim15$~kpc at the distance of DDO~44.

\begin{figure*}[!t]
\begin{center}
\includegraphics[width=0.975\textwidth, trim=1.0in 0.0in 1.5in 0.5in, clip]{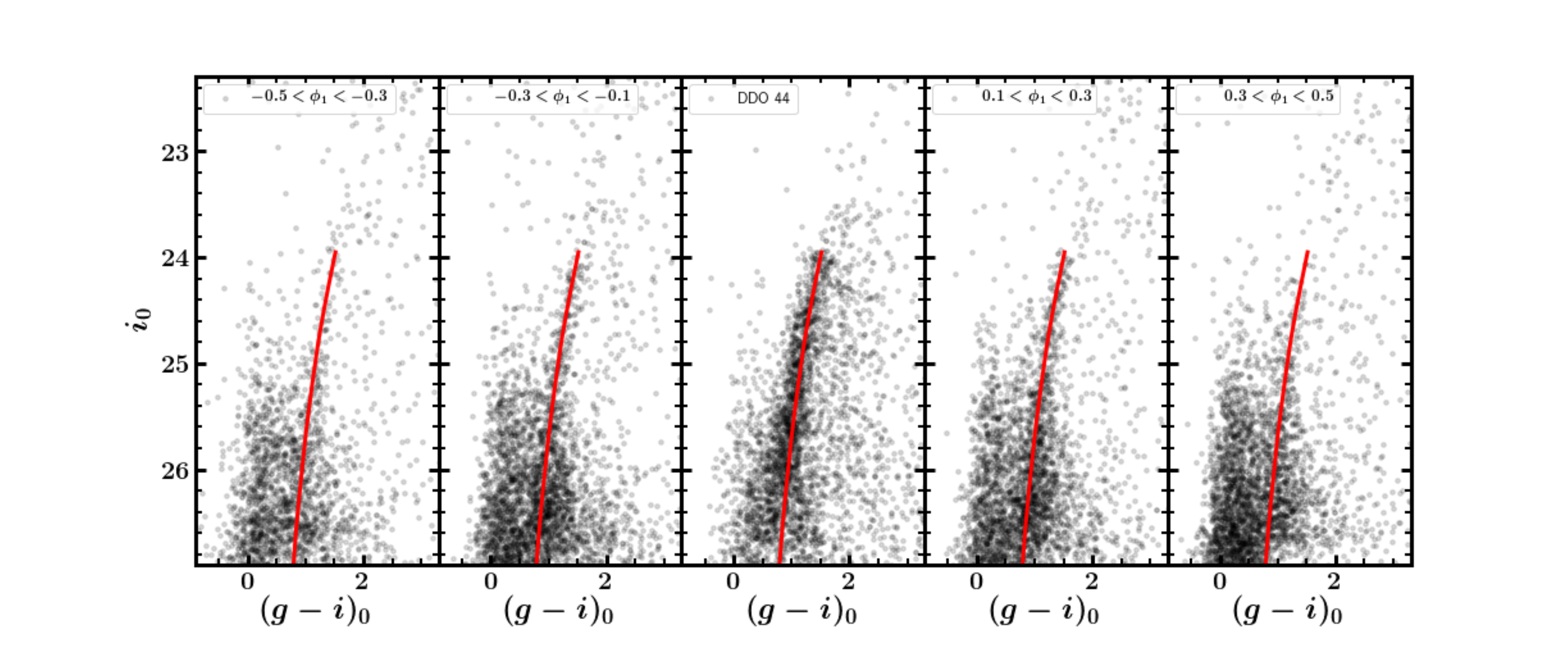}
\caption{CMDs selected in angular bins along the stream, within $\left|\phi_2\right| < 0.05\arcdeg$ of the stream center. The isochrone is the same [Fe/H] = $-1.6$ isochrone used in Figure~\ref{fig:cmd_ddo44}. Stellar populations consistent with DDO~44 are seen to at least $\left|\phi_1\right| = 0.3\arcdeg$ ($\sim16$~kpc at the distance of DDO~44), and perhaps as far as $\left|\phi_1\right| = 0.5\arcdeg$ ($\sim26$~kpc at the distance of DDO~44) from DDO~44.
\label{fig:stream_CMDS}}
\end{center}
\end{figure*}

Fig.~\ref{fig:stream_CMDS} highlights CMDs of stars extracted in bins along the stream. The central panel contains the core of DDO~44 (within $0.1^\circ$), with panels to the left (south) and right (north) showing similar stellar populations extending into the stream. On both sides of DDO~44, the stream is barely noticeable (if at all) at $\left|\phi_1\right| \gtrsim 0.4^\circ$. The best-fit isochrone with [Fe/H] = $-1.6$ is a good match to the RGB stellar populations in both the core and stream. The bright AGB stars visible in the central regions of DDO~44 are seen in small numbers at all radii (and indeed, the density profile seen in Fig.~\ref{fig:stream_density} suggests that the AGB stars extend as far as the RGB stars in the stream). The fact that the oldest RGB stars and the intermediate-age AGB populations are both extended suggests that DDO~44 had little to no population gradient in its core before being tidally disrupted (however, \citealt{Lianou2010} found that metal-rich populations are more centrally concentrated in DDO~44 than metal-poor stars). 

\subsection{Total luminosity}\label{sec:stream_lum}

To estimate the total luminosity of DDO~44, including stars in its streams, we summed the flux of all RGB stars brighter than $i_0 < 26.5$, applying a completeness correction to each star's flux based on the fits in Figure~\ref{fig:completeness}. We then corrected for the unmeasured luminosity below the cutoff magnitude using a Dartmouth isochrone \citep{Dotter2008} with [Fe/H]$=-1.6$, 10 Gyr age, and power-law luminosity function slope of $-1.5$.\footnote{We note that if instead we use a Salpeter IMF, our derived total luminosity of DDO~44 changes by $<0.03$ mags.} From this luminosity function, we determine that $\sim22\%$ of the flux is in stars brighter than $i = 26.5$; we thus apply a correction to the total flux to account for the remaining 78\% of the light. Finally, we account for the ``missing'' data due to stellar crowding near the center of DDO~44 by excluding the inner $2\arcmin$ from our calculations. Adopting $\mu_{\rm R, 0} = 24.1$~mag~arcsec$^{-2}$ and a scale length of $39\arcsec$ \citep{Karachentsev1999a}, we estimate that $\sim60\%$ of the light is contained within our excluded $2\arcmin$ region.

We find $M_{i, {\rm tot}} = -13.4$ and $M_{g, {\rm tot}} = -12.6$ for the total luminosity of DDO~44 and the stars in its streams, where we take the region between $|\phi_1| < 0.1\arcdeg$ to be the main body of DDO~44. Of the total, $\sim17\%$ and $\sim11\%$ of the flux is contained in the northern ($0.1\arcdeg < \phi_1 < 0.7\arcdeg$) and southern ($-0.7\arcdeg < \phi_1 < -0.1\arcdeg$) portions of the stream, respectively. 
Given the many large corrections detailed in the previous paragraph, it is difficult to place uncertainties on these estimates. To facilitate comparison to the measurement by \citet{Karachentsev1999a} of $M_{R} = -13.1$, we transform these absolute magnitudes in the PanSTARRS system to Johnson-Cousins $R$-band using the relations from Table~6 of \citet{Tonry2012}. This yields a total $M_{R} \sim -13.3$ based on our HSC measurements. Removing the $\sim20-30\%$ of the resolved stars' flux that we find beyond $6\arcmin$ of the DDO~44 center would reduce our derived luminosity of DDO~44 by $\sim0.2-0.3$ mag, placing our estimate for the central body of the galaxy in excellent agreement with that of \citet{Karachentsev1999a}.

To make connection with theory, it is useful to translate from the object's absolute magnitude to stellar mass.  Our measured luminosity transforms to $M_V = -12.9$ (it is Local Group convention to report $V$-band absolute magnitudes), which corresponds to $M_{*} = 2.0\times10^7~M_\odot$ (assuming $V$-band stellar $(M/L)_{\rm V}$ of 1.6; \citealt{Woo2008}). Note that an estimate based on the $K$-band luminosity from \citealt{Karachentsev2013}, assuming $(M/L)_{\rm K} = 1$,  gives $M_{*} = 6.0\times10^7~M_\odot$ for DDO~44.

By fitting Gaussians to the resolved stellar surface density along the major and minor axes, we find an ellipticity $\epsilon \equiv 1 - b/a \approx 0.6$. It is unsurprising to find that DDO~44 is rather extended, and that its ellipticity is similar to that of the tidally disrupting Sagittarius dSph ($\epsilon = 0.64$; \citealt{McConnachie2012}). Also like the Sagittarius dSph, DDO~44's surface brightness (as measured by \citealt{Jerjen2001}) lies below typical dwarfs at its luminosity (see, e.g., Figure~7 from \citealt{McConnachie2012}), as expected for a system undergoing tidal disruption.  DDO~44's closest analogs in luminosity and stellar mass are, according to \citet{McConnachie2012}, Sagittarius ($M_V = -13.5$), Fornax ($M_V = -13.4$), and  And~VII ($M_V = -12.6$).  Of these three, only Sagittarius is clearly disrupting --- deep imaging data show no hints of tidal features for Fornax \citep{mywang2019}, and the ellipticities of Fornax and And~VII are far lower.  
Our derived metallicity for DDO~44 of [Fe/H] $= -1.6$ is near, but slightly on the low metallicity side of, the luminosity-metallicity relation for Local Group dwarf galaxies (e.g., \citealt{kirby2013a}, \citealt{McConnachie2012}). The metallicities of Sagittarius, Fornax, and And~VII are all significantly higher \citep{Kalirai2010,Kirby2011,Carlin2012,McConnachie2012,Hasselquist2019}.

We summarize the properties of DDO~44 and its stream in Table~\ref{tab:props}, including some relevant data from the literature.

\section{DDO 44 and its streams in context}\label{sec:context}

DDO~44 is clearly a disrupting dwarf, but questions remain about its history and association with a larger host.  In this section, we argue that NGC~2403 is the most likely host for DDO~44.  This conclusion allows us to use simulations to estimate how rare (or not) it is for a large dwarf to be disrupting around a low-mass host, and consider how the orbit of DDO~44 explains various features of its star-formation history (SFH).  We may also place DDO~44 in the context of the NGC~2403 satellite system, and consider whether NGC~2403's satellite luminosity function is in line with expectations from the $\Lambda$CDM paradigm.

\subsection{DDO~44 is a satellite of NGC~2403}\label{sec:sat_of_2403}

We consider whether DDO~44 is in fact a satellite of NGC~2403 or of the neighboring galaxy NGC~2366.

DDO~44 has a heliocentric radial velocity of $213~{\rm km~s}^{-1}$ \citep{Karachentsev2011,Tully2016}\footnote{Note that this velocity is apparently based solely on an HII region offset from the center of DDO~44, but likely associated with it. We could not locate any extant velocity measurements based on the stellar body of DDO~44.}, while the Tully~et~al. ``COSMICFLOWS-3'' catalog gives $v_{\rm hel} = 141~{\rm km~s}^{-1}$ for NGC~2403. This small difference in their relative velocities (for context, this velocity difference of $\sim70$~km~s$^{-1}$ is much less than the escape velocity from NGC~2403 of $\gtrsim200$~km~s$^{-1}$; see Fig.~\ref{fig:vr_plot}), in addition to the nearly identical distance moduli of DDO~44 and NGC~2403 (Sec.~\ref{sec:dist_isofit}) is suggestive of an association between the two galaxies.  

\begin{table}[!t] 
\renewcommand{\thetable}{\arabic{table}}
\centering
\small
\caption{Properties of DDO~44 and its stream} \label{tab:props}
\begin{tabular}{lcc}
\tablewidth{0pt}
\hline
\hline
Parameter & Value & Reference \\
\hline
RA & 07:34:11.50 & NED\\
Dec & +66:52:47.0 & NED\\
$m-M$ & 27.36$\pm$0.07 & this work\\
$D$ (Mpc) & 2.96$\pm$0.10 & this work\\
$M_{\rm V}$ & $-12.9$ & this work\\
$M_{\ast}$ ($M_{\odot}$) & $2\times10^7$ & this work\\
$\left[ {\rm Fe/H} \right]$ & $-1.6\pm0.3$ & this work\\
Ellipticity & 0.6 & this work\\
$R_{\rm e}$ (kpc) & 0.74$\pm$0.02 & \citet{Jerjen2001}\\
$<\mu_e>$ (B) & 26.00 & \citet{Jerjen2001}\\
H\textsc{i} mass ($M_{\odot}$) & $<10^6$  & KK07\tablenotemark{a}\\
stream extent ($\arcdeg$) & $\sim1\arcdeg$ & this work\\
stream extent (kpc) & $\sim50$ & this work\\
$L_{\rm stream}/L_{\rm tot}$\tablenotemark{b} & $\sim20-30\%$ & this work\\

\hline
\end{tabular}
\tablenotetext{a}{\citet{Karachentsev2007}}
\tablenotetext{b}{Fraction of luminosity in the stream.}
\end{table}

We also note that the extension of the DDO~44 stream points in the general direction of nearby, roughly SMC stellar-mass, galaxy NGC~2366 ($D \sim 3.3$~Mpc; \citealt{Karachentsev2013}), which is $\sim2.4\arcdeg$ north (position angle $348\arcdeg$ east of north) of DDO~44. In spite of the fact DDO~44's projected distance from NGC~2366 of $\sim130$~kpc is likely beyond the virial radius of NGC~2366 ($R_{\rm vir} \sim 110$~kpc for a slightly sub-SMC stellar-mass galaxy; e.g., \citealt{Dooley2017b}), it is interesting to note that its radial velocity of 103~km~s$^{-1}$ is similar to the 213~km~s$^{-1}$ velocity of DDO~44 (i.e., the difference of $\sim110$~km~s$^{-1}$ is likely less than the escape velocity of NGC~2366). NGC~2366 is actively star forming, with distortions in its H\textsc{i} contours \citep{Lelli2014}, indicating a recent interaction (which Lelli et al. attributed to an ongoing minor merger with NGC~2363). We believe it is unlikely that NGC~2366 caused the visible damage to DDO~44. NGC~2366 is a gas-rich dwarf galaxy, with ordered rotation and a rotation velocity of only $\sim60$~km~s$^{-1}$ \citep{Oh2008}. Only a small fraction of its H\textsc{i} layer is discrepant from the otherwise well-behaved rotation. In order to feel significant tidal effects, DDO~44 would have needed to pass very near NGC~2366, in which case NGC~2366 would have felt strong tidal forces in this $\sim25:1$ stellar mass-ratio interaction (examples of similar mass-ratio interactions in the Local Volume can be seen in \citealt{Pearson2016}). It thus seems implausible that a galaxy as small as NGC~2366 has stripped away all the gas, $\sim90\%$ of the dark matter, and $\sim25\%$ of the stars in DDO~44.

In summary, many pieces of evidence make it more likely that DDO~44 interacted with the more massive NGC~2403 than its less massive neighbor NGC~2366: (1) DDO~44 is closer to NGC~2403 in projected and line-of-sight separation than it is to NGC~2366, (2) DDO~44 is likely beyond the virial radius of NGC~2366 (and within that of NGC~2403), (3) DDO~44's velocity is closer to that of NGC~2403 than to NGC~2366, (4) this relative velocity is less than the expected escape velocity of NGC~2403, and (5) NGC~2403 is more massive than NGC~2366, and thus more likely to host (and retain) a large dSph such as DDO~44.

\subsection{DDO~44 has an unusual orbit about its host}

In order to investigate how common disrupted satellites like DDO 44 are around galaxies like NGC 2403, we search for analog systems in cosmological simulations. If analog systems are common, we can use the present-day kinematics of the DDO 44 analogs to estimate pericenter and apocenter distributions and use merger trees to track their orbital histories, giving us insight into the interaction history of the system. If very few analog systems are identified, then we can infer that the DDO 44 -- NGC 2403 system is, in some way, an extreme interaction. We may also use the infall time distribution function and the orbital history to constrain models for the physical origin of the truncation of DDO~44's SFH.  This analysis is based in part on work by \cite{Rocha2012}, and is similar to the approach taken by \citet{Besla2018}. A full description is found in \citet{Garling2019}, but a brief description is given below. 

\begin{figure*}[th]
\includegraphics[width=0.49\textwidth,page=1]{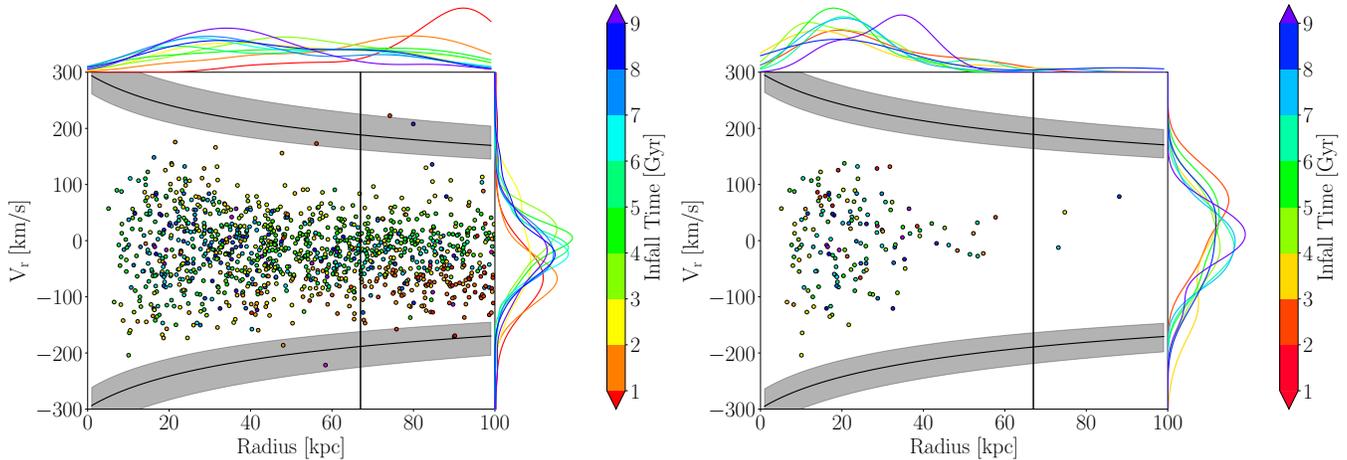}
\includegraphics[width=0.49\textwidth,page=2]{analysis_firstinfall_100}
\caption{{\it Left panel:} Distribution of DDO 44 analogs from Illustris-1 in radial velocity and radius (relative to their host), without the tidal disruption criterion applied. Typical escape velocity curves are overplotted, and the projected distance of DDO 44 from NGC 2403 (67 kpc) is shown with a solid line. Illustris-1 analogs are color-coded according to infall time, with the most recent infalls labeled in red and the earliest infalls labeled purple.  The relative host-subhalo radial velocity (left) and separation (top) probability distribution functions are given as a function of infall time.  {\it Right panel:} Same as left, but with the tidal disruption criterion of $M_{z=0} / M_{\text{infall}} \leq 0.1$ applied. This population consists of primarily early accretions which have made several orbits, and only three systems have 3D radii that are greater than DDO 44's projected radius.
\label{fig:vr_plot}}
\end{figure*}

We use the Illustris simulations \citep{Vogelsberger2013,Genel2014,Vogelsberger2014a,Vogelsberger2014b,Nelson2015,Rodriguez-Gomez2015} for this analysis. Illustris is simulated with WMAP9 $\Lambda$CDM cosmological parameters \citep{Hinshaw2013}. We use the flagship Illustris-1 run, which simulates a comoving box of volume 106.5 Mpc\textsuperscript{3} with $1820^3$ particles each of dark matter, gas, and tracers that are used to track the Lagrangian evolution of the gas \citep{Genel2013}. We replicate our analysis on the Illustris-1-Dark run, which simulates the same volume as Illustris-1 with the same number of dark matter particles but without hydrodynamics, and find no significant differences from the analysis presented below for Illustris-1. We utilize the friends-of-friends group catalogs, \textsc{subfind} subhalo catalogs, and the \textsc{sublink} merger trees \citep{Rodriguez-Gomez2015} to identify halos and subhalos and track their evolution through time.

Because the hydrodynamic mass resolution is too low for us to identify analog systems using the simulated stellar masses, we instead find analog systems based on dark-matter halo masses and abundance matching.  We identify analog systems in Illustris by matching \textsc{subfind} subhalo masses to halo masses for DDO 44 and NGC 2403, calculated by converting their stellar masses to halo masses using the abundance matching scheme of \cite{Moster2013}, which returns halo masses of $2.8 \times 10^{10} M_{\odot}$ and $2.5 \times 10^{11} M_{\odot}$ for DDO~44 and NGC~2403, respectively. 

In searching for analog systems, we require that NGC 2403 analogs be the most massive subhalo in their group (i.e., the host) and DDO 44 analogs be the second most massive subhalo in the group (i.e., most massive satellite). To take into account scatter in the stellar-to-halo-mass relation and the uncertainty in stellar mass, we accept systems that have masses within a factor of two of the above halo masses. For the DDO 44 analogs, we place this mass constraint on first infall mass rather than present day mass, as the majority of a satellite's stellar mass is typically formed prior to accretion by its present day host (i.e., when it was a central galaxy of its own), and we expect mass loss to be significant for true DDO 44 analogs. We define first infall as the time when DDO 44 analogs first enter the virial radius of their present-day NGC 2403 analog hosts.

Given that DDO 44's stellar population is disrupted, we can infer that its dark matter halo is as well. Because dwarf stellar populations are deeply embedded within their dark matter halos, the degree of disruption to the dark matter halo must be severe --- \cite{Penarrubia2008} showed that King profiles embedded in NFW halos must have $\sim 90\%$ of their dark matter halo stripped before stars begin to be disrupted. This criterion is valid regardless of whether the subhalo is cored or cusped, as differences in tidal stripping between the two density profiles only become important when the tidal radius approaches the size of the core, which typically has an enclosed mass less than 10\% of the infall halo mass \citep{Dooley2016,GK2017}.  As such, we place an additional disruption criterion on our DDO 44 analogs and only accept systems with $M_{z=0} / M_{\text{infall}} \leq 0.1$. This proves to be a very strong condition, as shown in Fig.~\ref{fig:vr_plot}. There are 1628 undisrupted analog systems, but only 157 systems remain when the disruption criterion is imposed.\footnote{Note that systems for which the halo finder does not identify a subhalo remnant are not included in this count. Systems that may be missed by the halo finder include satellites that are fully disrupted or where the remaining mass lies below the resolution limit of the simulation, or where a stripped subhalo lies close to the host halo center even if the subhalo is not fully disrupted.} These remaining systems have predominantly early infall times (mean $\sim$ 7 Gyr), have made several complete orbits with present-day mean eccentricity $\sim 0.5$, and have small apocenters of 30--70 kpc; we interpret the majority of this population to represent satellites that were accreted early and lost mass gradually. Given DDO 44's projected distance of 67 kpc (and thus likely a much larger apocenter), we do not believe this population is representative of DDO 44.

Next we discuss the three systems with radii greater than 67 kpc, shown to the right of the line in Fig.~\ref{fig:vr_plot}. Two of these systems have relatively early infall times (7--9 Gyr). These systems have much larger apocenters than is typical for other satellites in the population with similar infall times ($>150$ kpc compared to 30--70 kpc). We attribute this to their more eccentric orbits ($e>0.7$ at present-day) which allow them to maintain low binding energies over many orbits due to the lessened effects of dynamical friction compared to the bulk of the population.  For reasons associated with the SFH and the morphology of NGC~2403's gas, described in Sec. \ref{sec:sfh}, we consider these orbits unlikely.

Most interesting is the subhalo identified with an infall time $\sim 3.5$ Gyr and an apocenter of $\sim150$ kpc, much larger than DDO 44's projected distance of 67 kpc. This subhalo has an orbital eccentricity greater than 0.8 at present-day, and has lost more than 90\% of its mass over only a handful of orbits. 
Such a configuration, with a highly radial orbit, would be highly effective at stripping the dark matter halo sufficiently to disrupt the stellar population. 

In summary, we find that DDO~44's existence is rare for a host galaxy like NGC~2403.  Limiting our simulated sample to NGC~2403-mass analogs with DDO~44-mass analog satellites, we find that only $\sim 0.1\%$ of these analogs have a disrupting DDO~44 with a large separation between host and satellite.  About 10\% of analogs are disrupting but are located much closer to the host.  We argue that DDO~44 is likely a recently accreted satellite with a highly eccentric orbit.

\subsection{Insights from DDO~44's orbit on its star-formation history}\label{sec:sfh}

We now combine the orbital information from the previous section with DDO~44's current gas content and measured SFH to estimate a timescale and identify the physical process likely to be responsible for the end of star formation in this galaxy.

The SFH of DDO~44 \citep{Girardi2010,Weisz2011} suggests that it formed $\sim15-20\%$ of its stars between 1--3~Gyr ago. However, \citet{Karachentsev2007} found an upper limit on the neutral hydrogen content of DDO~44 of only $M_{\rm HI} < 10^6 M_\odot$, and no $H\alpha$ emission associated with DDO~44.\footnote{We note that \citet{Karachentsev2011} found evidence of a small ($\sim4\arcsec$ in size), off-center H\textsc{ii} region possibly associated with DDO~44, with H$\alpha$ emission and possibly some late B-type stars associated with it. \citet{Karachentsev2011} suggest that this small bit of star formation is in a clump of accreted intergalactic gas. It is also possible that it was picked up during DDO~44's recent interaction with NGC~2403.} Given that it has no hint of gas, this star formation event (and/or the interaction leading to the tidal features) must have exhausted whatever neutral hydrogen DDO~44 still had. 

In this regard, it is interesting to note that \citet{DeBlok2014} suggested that a fly-by encounter between NGC~2403 and DDO~44 could be responsible for the anomalous cloud of H\textsc{i} detected at the northwest edge of the NGC~2403 disk. In their Fig.~8, \citet{DeBlok2014} overlay the contours of this H\textsc{i} cloud on a map of RGB stars from \citet{Barker2012}, showing a disturbance in the outer stellar disk (also visible in our RGB density map in Figure~\ref{fig:rgb_density}) at the location of the H\textsc{i} cloud, roughly in the direction of the DDO~44 stream's projected intersection with the NGC~2403 disk. The H\textsc{i} cloud has a mass of $\sim6\times10^6 M_\odot$. This is within roughly an order of magnitude of the stellar mass of DDO~44, so it is plausible that the H\textsc{i} cloud is material stripped from DDO~44 (though we note that the gas could also have been pulled from NGC~2403 during the interaction). Interestingly, \citet{Lianou2010} found that DDO 44 lies in the ``transition region'' in luminosity-metallicity space between the locations occupied by dSphs (i.e., typically quenched systems) and dIrrs (typically gas-rich, star-forming systems). Unlike the three other M81-group dSphs found by Lianou et al. to reside in this transition region, DDO~44 does not have H\textsc{i} or H$\alpha$, suggesting that its gas may have been removed recently.

A recent infall is potentially also supported by the SFH of NGC~2403.  We note that there was also a slight upturn in the SFH of NGC~2403 within the past $\sim2$~Gyr \citep{Williams2013}, which could possibly arise due to an interaction with a satellite such as DDO~44. 
We do note, however, that in the same study based on deep \textit{HST+ACS} observations \citep{Williams2013}, it is claimed that the disk of NGC~2403 appears ``remarkably undisturbed.'' Refined spectroscopic measurements of the velocity of DDO~44 (based on the stellar light rather than the lone H\textsc{ii} region) may enable modeling of its orbit that can reconcile the apparently conflicting information given by the H\textsc{i} and the stellar disk of NGC~2403.

These lines of evidence suggest a recent (1--2 Gyr ago) close interaction between DDO~44 and NGC~2403 that tidally stripped DDO~44's H\textsc{i} reservoir.  The travel time between the small pericenter and DDO~44's current position with respect to NGC~2403 is approximately 1 Gyr.  Because the gas scale-length generically exceeds the optical size of galaxies, stellar stripping is a sign that gas ought to have been heavily tidally stripped as well \citep{leisman2017}. While it is possible that gas may have additionally been ram-pressure stripped from DDO~44, the small pericenter implied by the H\textsc{i} distribution and SFH of NGC~2403 suggests tides play the dominant role in removing gas and quenching star formation in DDO~44.

In short, the SFH suggests a relatively recent infall for DDO~44, on a highly unusual orbit, for which much of the cold gas fuel for star formation was tidally stripped during the last pericenter passage.

\subsection{NGC~2403's satellite luminosity function}

We place DDO~44 in the context of its role as the most massive satellite of NGC~2403.  DDO~44 is one of two known satellites of NGC~2403, and is by far the most massive.  The stellar mass ratio between NGC~2403 ($M_* \sim 7\times10^9~M_\odot$) and DDO~44 is $\sim350$ (i.e., a stellar mass gap of $\Delta M_{*} = \log{(M_{\rm N2403}/M_{\rm max, sat})} \sim 2.5$). Because of the steepness of the halo mass function and the stellar-mass--halo-mass relation, this mass ratio is not unusual.  According to the analysis of satellite galaxies in the Sloan Digital Sky Survey \citep{York2000} by  \citet{Sales2013}, we expect there to be approximately one satellite with a stellar mass exceeding 0.1\% of the host's stellar mass in the virial volume of NGC~2403.  However, the mass gap between NGC~2403's first- and second-most-massive satellite is unusual. The gap between the stellar masses of DDO~44 and second-most-massive NGC~2403 satellite (MADCASH~J074238+652501-dw; \citealt{Carlin2016}; $M_{*} \sim 1\times10^5~M_\odot$) is a factor of $\sim200$.  Based on predictions from models (\citealt{Dooley2017b}; see also \citealt{Jahn2019} for predictions from the FIRE simulations), we expect that NGC~2403 should host between 2--8 satellites with stellar masses $>10^5 M_\odot$ (the uncertainty arises both from halo-to-halo variation and the as-yet poorly constrained stellar-mass--halo-mass relation on these scales), so the observation of only one satellite below DDO~44's stellar mass down to $M_*\sim 10^5 M_\odot$ is unusual in both the size of the gap and the total number of satellites.

While a systematic search and characterization of our completeness will be the subject of a future contribution, the fact that we only find two companions to NGC~2403 in our preliminary search implies that this LMC analog may be lacking bright satellites (relative to predictions).

\section{Discussion and Conclusions}\label{sec:discussion}

We report the discovery of a stellar tidal stream around the Local Volume dwarf spheroidal galaxy DDO~44, based on deep, resolved-star observations with Subaru+HSC. The tidal stream stretches $\sim25$~kpc on either side of the main body of DDO~44, and is oriented toward NGC~2403, of which DDO~44 is likely a satellite. We reconstruct the total luminosity of the DDO~44 progenitor, and find that it had a luminosity of at least $M_{i, \rm{tot}} = -13.4$ ($M_{g, \rm{tot}} = -12.6$; or $M_{\rm{V, tot}} = -12.9$). 

Using the Illustris simulation suite, we show that DDO~44 is an unusual object.  While disruptions by LMC-mass hosts are not uncommon (10\% of our mass-matched analog systems show a massive disrupting dwarf), and observed in other LMC analog systems \citep[notably NGC~4449;][]{Karachentsev2007_N4449, MD2012, Rich2012, Toloba2016}, the typical separation with respect to the host in Illustris is much smaller than observed for the NGC~2403-DDO~44 system.  We find only $\sim 0.1\%$ of our mass-matched analog systems are disrupting with the type of large observed separation between NGC~2403 and DDO~44.  Combined with observations of the H\textsc{i} distribution of both galaxies, the recent upturn of star-formation in NGC~2403, and the recent quenching of DDO~44, we argue that DDO~44 only recently entered the halo of NGC~2403 on a high-eccentricity orbit with a pericenter small enough to tidally strip both its stars and its gas reservoir. 

This work strengthens the case for significant interaction in dwarf pairs, even for dwarf systems with high mass ratios ($\sim 100$) like the NGC~2403-DDO~44 system.  NGC~2403's recent upturn in star-formation rate is consistent with our estimated infall and pericenter passage of DDO~44.  The TiNy Titans survey found that starbursts are more frequent for dwarf pairs than for field dwarfs, although they included systems with much lower mass ratios than we consider in this study \citep{Stierwalt2015}.  Furthermore, the quenching and obvious tidal stripping of DDO~44 shows that even low-mass hosts may exert considerable environmental influence over their satellites.  Although a systematic analysis of the strength and prevalence of environmental quenching of satellite galaxies by LMC-like hosts is beyond the scope of this work, we are revealing strong evidence that even low-mass hosts can quickly quench and destroy massive satellites.

\acknowledgments

We thank the referee for insightful comments that helped clarify this work. We acknowledge support from the following NSF grants: AST-1816196 (JLC); AST-1814208 (DC); AST-1813628 (AHGP and CTG); and AST-1616710 (AJR). AJR was supported as a Research Corporation for Science Advancement Cottrell Scholar. JS acknowledges support from the Packard Foundation. Research by DJS is supported by NSF grants AST-1821987, AST-1821967, AST-1813708, and AST-1813466. The work of authors JLC, DJS, and JRH was performed in part at the Aspen Center for Physics, which is supported by National Science Foundation grant PHY-1607611.

This research has made use of NASA's Astrophysics Data System, and \texttt{Astropy}, a community-developed core Python package for Astronomy \citep{Price-Whelan2018b}. 

The Pan-STARRS1 Surveys have been made possible through contributions of the Institute for Astronomy, the University of Hawaii, the Pan-STARRS Project Office, the Max-Planck Society and its participating institutes, the Max Planck Institute for Astronomy, Heidelberg and the Max Planck Institute for Extraterrestrial Physics, Garching, The Johns Hopkins University, Durham University, the University of Edinburgh, Queen's University Belfast, the Harvard-Smithsonian Center for Astrophysics, the Las Cumbres Observatory Global Telescope Network Incorporated, the National Central University of Taiwan, the Space Telescope Science Institute, the National Aeronautics and Space Administration under Grant No. NNX08AR22G issued through the Planetary Science Division of the NASA Science Mission Directorate, the National Science Foundation under Grant AST-1238877, the University of Maryland, Eotvos Lorand University (ELTE), and the Los Alamos National Laboratory.

\vspace{5mm}
\facilities{Subaru+HSC, PS1}

\software{\texttt{astropy} \citep{TheAstropyCollaboration2013,Price-Whelan2018b}, \texttt{gala} \citep{Price-Whelan2017}, \texttt{Matplotlib} \citep{Hunter2007}, \texttt{NumPy} \citep{VanderWalt2011}, \texttt{Topcat} \citep{Taylor2005}.}

\bibliographystyle{aasjournal}

\end{document}